\documentclass[structabstract, letter, longauth]{aa}
\usepackage{graphicx}
\usepackage{txfonts}
\begin{document}
   \title{First detection of the Methylidyne cation (CH$^+$) fundamental rotational line with the \textit{Herschel}/SPIRE FTS
   \thanks{\textit{Herschel} is an ESA space observatory with science instruments provided
by European-led Principal Investigator consortia and with important participation from NASA.}}

\author{
D. A. Naylor\inst{1}, E. Dartois\inst{2},  E. Habart\inst{2}, A. Abergel\inst{2}, J.-P. Baluteau\inst{3}, S.C. Jones $^{1}$, E. Polehampton$^{1,4}$,
P. Ade\inst{5}, L. D. Anderson\inst{3}, P. Andr\'e\inst{6}, H. Arab\inst{2},  J.-P. Bernard\inst{7}, K. Blagrave\inst{8},
F. Boulanger\inst{2}, M. Cohen\inst{9}, M. Compi\`{e}gne\inst{8}, P. Cox\inst{10}, G. Davis\inst{11}, R. Emery\inst{4}, T. Fulton\inst{12},
C. Gry\inst{3}, M. Huang\inst{13}, C. Joblin\inst{8,17}, J. M. Kirk\inst{5},  G. Lagache\inst{2}, T. Lim\inst{4},
S. Madden\inst{6}, G. Makiwa\inst{1}, P. Martin\inst{9}, M.-A. Miville-Desch\^enes\inst{2},
S. Molinari\inst{14}, H. Moseley$^{15}$, F. Motte\inst{6}, K. Okumura\inst{6},
D. Pinheiro Gocalvez\inst{9}, J. A. Rod\'on\inst{3}, D. Russeil\inst{3}, P. Saraceno\inst{14}, S. Sidher\inst{4}, L. Spencer\inst{1}, B. Swinyard\inst{4},
D. Ward-Thompson\inst{5}, G. J. White\inst{4,16},  A. Zavagno\inst{3}
  }
 \institute{
 Institute for Space Imaging Science, University of Lethbridge, Alberta, Canada \and
 Institut d'Astrophysique Spatiale, CNRS/Universit\'e Paris-Sud 11, 91405 Orsay, France
 \and
 Laboratoire d'Astrophysique de Marseille, UMR6110 CNRS, 38 rue F.
Joliot-Curie, F-13388 Marseille France \and
 Space Science Department, Rutherford Appleton Laboratory, Chilton, UK \and
 Department of Physics and Astronomy, Cardiff University, Cardiff, UK \and
 CEA, Laboratoire AIM, Irfu/SAp, Orme des Merisiers, F-91191Gif-sur-Yvette, France \and
 Centre d'Etude Spatiale des Rayonnements, CNRS/Universit\'e de Toulouse, 9 Avenue du colonel Roche, BP 44346, 31028 Toulouse Cedex 04, France \and
 Canadian Institute for Theoretical Astrophysics, Toronto, Ontario, M5S 3H8, Canada  \and
 University of California, Radio Astronomy Laboratory, Berkeley, 601 Campbell Hall, US Berkeley CA 94720-3411, USA \and
Institut de Radioastronomie Millim\'etrique (IRAM), 300 rue de la Piscine, F-38406 Saint Martin dÕH\`eres, France  \and
Joint Astronomy Centre, University Park, Hilo, USA \and
Blue Sky Spectroscopy Inc, Lethbridge, Canada \and
National Astronomical Observatories (China) \and
 Istituto di Fisica dello Spazio Interplanetario, INAF, Via del Fosso
del Cavaliere 100, I-00133 Roma, Italy \and
 NASA - Goddard SFC, USA \and
 Department of Physics and Astronomy, The Open University, Milton Keynes MK7 6AA, UK \and
CNRS, UMR5187, 31028 Toulouse, France
%
           }

   \date{Received March 31, 2010; accepted May 4, 2010}


  \abstract
   {}
   {To follow the species chemistry arising in diverse sources of the Galaxy with \textit{Herschel}.}
   {SPIRE FTS sparse sampled maps of the Orion bar \& compact HII regions G29.96-0.02 and G32.80+0.19 have been analyzed.}
   {Beyond the wealth of atomic and molecular lines detected in the high-resolution spectra obtained with the FTS of SPIRE in the Orion Bar, one emission line is found to lie at the position of the fundamental rotational transition of CH$^+$ as measured precisely in the laboratory (Pearson \& Drouion 2006). This coincidence suggests that it is the first detection of the fundamental rotational transition of CH$^+$. This claim is strengthened by the observation of the lambda doublet transitions arising from its relative, CH, which are also observed in the same spectrum. The broad spectral coverage of the SPIRE FTS allows for the simultaneous measurement of these closely related chemically species, under the same observing conditions. The importance of these lines are discussed and a comparison with results obtained from models of the Photon Dominated Region (PDR) of Orion are presented. The CH$^+$ line also appears in absorption in the spectra of the two galactic compact HII regions G29.96-0.02 and G32.80+0.19, which is likely due to the presence of CH$^+$ in the the Cold Neutral Medium of the galactic plane. These detections will shed light on the formation processes and on the existence of CH$^+$, which are still outstanding questions in astrophysics. }
   {}
   \keywords{\textit{Herschel}, SPIRE, FTS, methylidyne, CH$^+$, CH, Orion, G29.96-0.02, G32.80+0.19}
   \authorrunning{Naylor et al.}
   \maketitle

\section{Introduction}
Molecules and radicals with C and H as constituents are expected to be the first chemical building blocks.
The  methylidyne cation (CH$^+$) was discovered at visible wavelengths 70 years ago by \cite{Douglas1941}, shortly after the discovery of the methylidyne (CH) radical by \cite{Swings1937}. Since that time there has been intense debate surrounding the chemistry of this species (Bates and Spitzer 1951). CH$^+$ far-IR detections were reported by \cite{Cernicharo1997} for the J=2-1 to 4-3 transitions in the NGC7027 PDR using Infrared Space Observatory spectra. \cite{Falgarone2005} report the probable detection of $\rm^{13}CH^+$(1-0) in absorption against a galactic source.
CH N=2-1 transitions have been detected by \cite{Stacey1987} and \cite{Polehampton2005} toward  SgrB2 and by \cite{Liu1997} in NGC7027.
CH$^+$ is commonly detected in the visible and found to correlate with rotationally excited H$_2$ (Lambert and Danks 1986), which argues for its formation in energetic regions. One of the underlying questions is whether CH$^+$ formation proceeds via a still undefined reaction or, following the consensus in the literature, via ISM conditions liberating sufficient energy (shock chemistry, turbulence, UV pumping) to overcome activation barriers, thereby strongly enhancing the otherwise hindered endothermic reaction rate for $\rm C^+ + H_2 \rightarrow CH^+ + H$, with a $\sim$ 0.4eV barrier.
CH$^+$ formation has been evoked as a result of shock chemistry (Pineau des Forets et al. 1986, Draine and Katz 1986 and citations). Species produced in shocks, however, should display velocity differences from the source producing the shocks. The CH$^+$ absorption profiles measured to date are found to lie at the same velocity as CH (\cite{Gredel1997}), which has been taken as an argument against shock chemistry production.
UV pumping, leading to vibrationally excited H$_2$ (Stecher and Williams 1972), is not expected to be significant in the diffuse ISM, on the grounds that the mean UV field is not strong enough to generate the observed excited H$_2$. However, the laboratory measured $\rm C^+ + H_2(v=1) \rightarrow CH^+ + H$ reaction (Hierl et al. 1997), may support such a route to CH$^+$ formation in PDRs (e.g. Agundez et al. 2010).
Among the propositions, turbulence models are gaining support in CH$^+$ chemistry (Godard et al. 2009 and references therein).
In this letter we present the first detection of the fundamental rotational transition of CH$^+$, along several lines-of-sight, in different and distinct regimes of the ISM and discuss the potential of SPIRE FTS observations to constrain the CH chemistry in such sources.
%
%
\begin{table}

\caption{Observations summary}             
\label{table:1}      
\begin{center}                    
\tiny
\begin{tabular}{l  l l l l c}        
\hline\hline                 
Source & Obs. date &Ra &Dec &Int(s) \\    
\hline                        
HD37041 		 	&$09/09/13$ 	&$\rm 05h35m22.83s$ 	&$\rm {-}05d24'57.67''$ &$266.4$\\
G29.96$ -$0.02	 	&$09/09/13$ 	&$\rm 18h46m04.07s$ 	&$\rm {-}02d39'21.88''$ &$266.4$\\
G32.80+0.19	 	&$09/09/21$ 	&$\rm 18h50m30.87s$ 	&$\rm 00d02'00.85''$ &$266.4$\\
\hline                                   
\end{tabular}
\end{center}
\vspace*{-0.5cm}
\end{table}
\section{Observations with the SPIRE FTS}
\begin{figure}[htbp]
\begin{center}
\includegraphics[width=0.9\columnwidth]{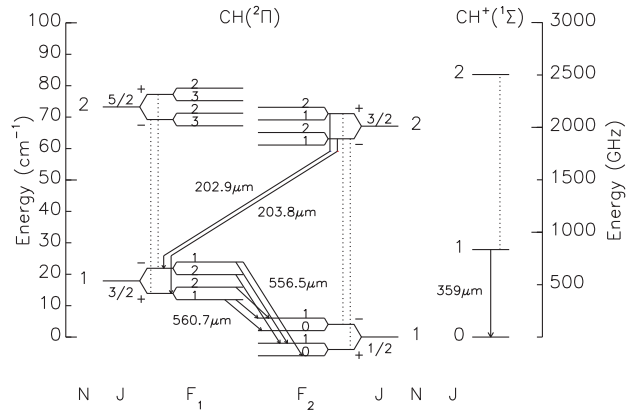}
\caption{Energy diagram for CH and CH$^+$ in the SPIRE FTS range. Expected transitions are shown with arrows. Dotted lines are transitions lying at higher frequencies accessible by other \textit{Herschel} instruments.}
\label{diagram}
\end{center}
\vspace*{-0.5cm}
\end{figure}

The SPIRE FTS measures, simultaneously, the source spectrum across two wavebands:  SLW, covering 14.9 - 33.0 cm$^{-1}$ (303-671 $\mu$m) and SSW covering 32.0-51.5 cm$^{-1}$ (194-313 $\mu$m). Each band is imaged with a hexagonal bolometer array with pixel spacing of approximately twice the beam-width. The FWHM beam-widths of the SLW and SSW arrays vary between  29-42" and 17-21" respectively. The source spectrum, including the continuum, is obtained by taking the inverse transform of the observed interferogram. For more details on the ESA \textit{Herschel} Space Observatory, the SPIRE FTS, and the FTS calibration and data reduction procedures, the reader is referred to the articles by Pilbratt et al. (2010), Griffin et al. (2010), and Swinyard et al. (2010), respectively, in this volume.
Our observations are part of the {\emph{Evolution of Interstellar dust}} key project of the SPIRE consortium (\cite{Abergel2010}). The source data presented here include the Orion Bar as well as the G29.96$ -$0.02 and G32.80+0.19 compact HII regions. A short observation log is given in Table\ref{table:1}. For example, G29.96-0.02 was observed with the high-resolution mode of the SPIRE FTS on the 13th of September, 2009 at 19:49 (\textit{Herschel} observation ID, 1342183824). Two scan repetitions were observed giving an on-source integration time of 266.4 seconds. The unapodized spectral resolution was 0.04 cm$^{-1}$ (1.2 GHz). After apodization (using extended Norton-Beer function 1.5; Naylor and Tahic 2007) the FWHM of the resulting instrument line shape is 0.0724 cm$^{-1}$ (2.17 GHz).
While unapodized FTS spectra provide the highest spectral resolution, the instrument line shape, which in the case of an ideal FTS is the classical sinc function, is characterized by relatively large secondary oscillations having negative lobes. An iterative spectral line fitting routine has been developed to extract line parameters from unapodized FTS spectra. This algorithm fits a continuum (either a low order polynomial or a blackbody variant) and a series of lines using the Levenberg-Marquardt least squares method. The fitting procedure weights the spectral intensity at a given frequency of an averaged spectrum by the statistical uncertainty at that frequency. The fitting routine returns the line centres, intensities, and widths together with their associated errors.
The transitions of methylidyne and its cation accessible to the FTS are summarized in Fig~\ref{diagram} and Table~\ref{table:2}.
\subsection*{The Orion Bar}
\begin{figure*}[htbp]
\begin{minipage}{0.7\columnwidth}
\begin{center}
\hspace*{-0.8cm}
\includegraphics[width=0.9\columnwidth]{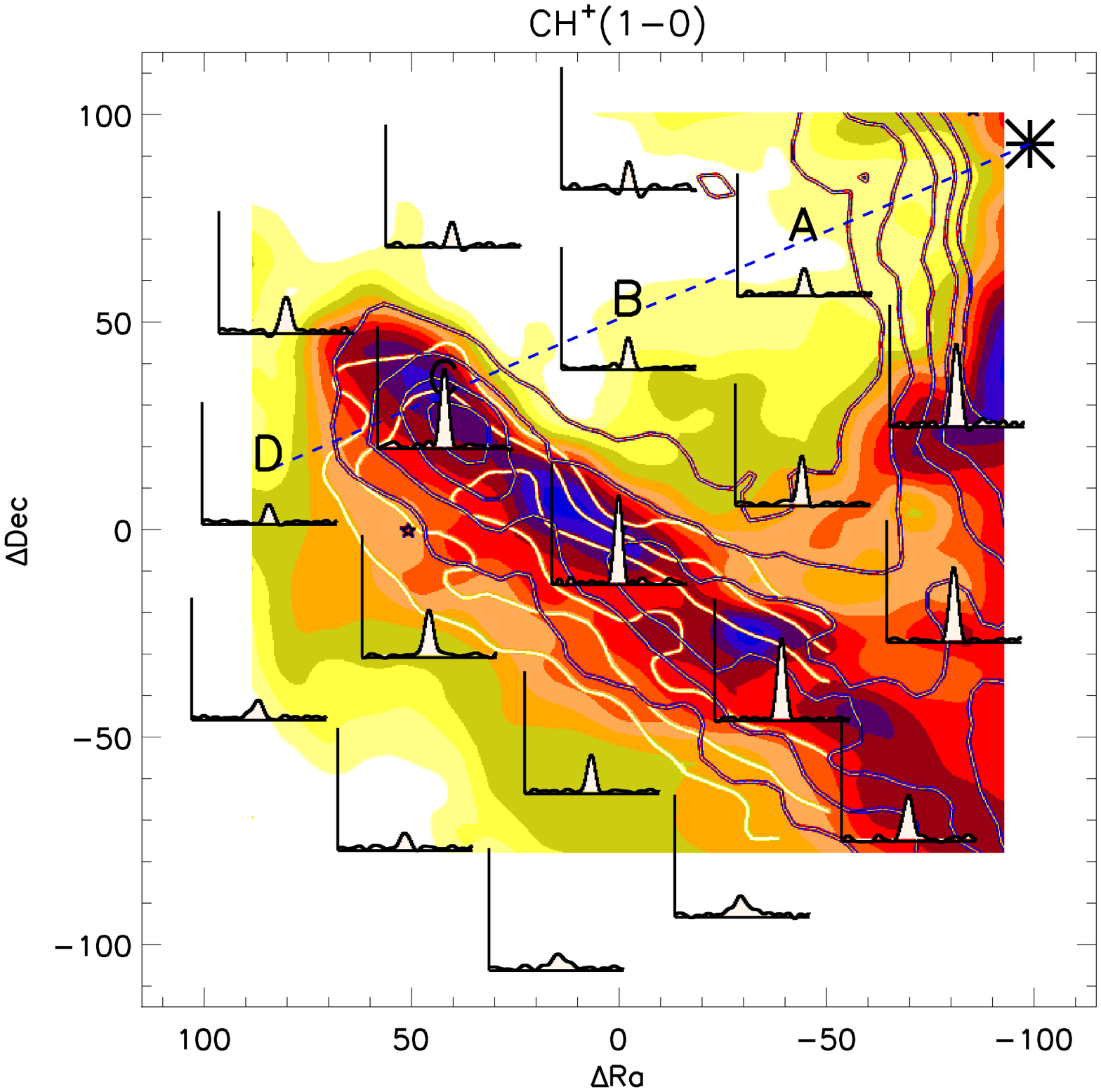}
\end{center}
\end{minipage}
\begin{minipage}{0.7\columnwidth}
\begin{center}
\hspace*{-1cm}
\includegraphics[width=0.9\columnwidth]{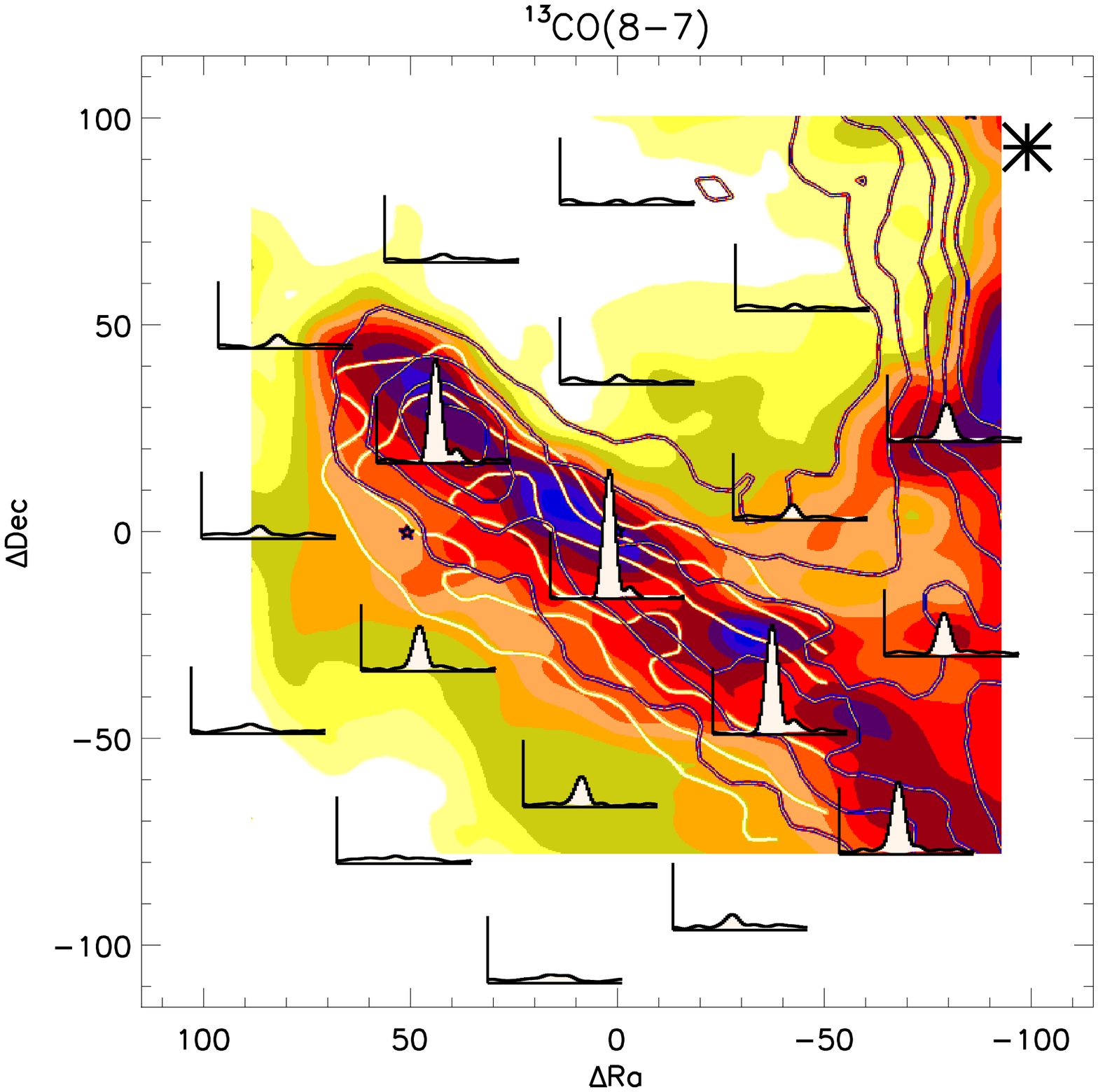}
\end{center}
\end{minipage}
\begin{minipage}{0.7\columnwidth}
\begin{center}
\hspace*{-0.5cm}
   \includegraphics[width=0.9\columnwidth]{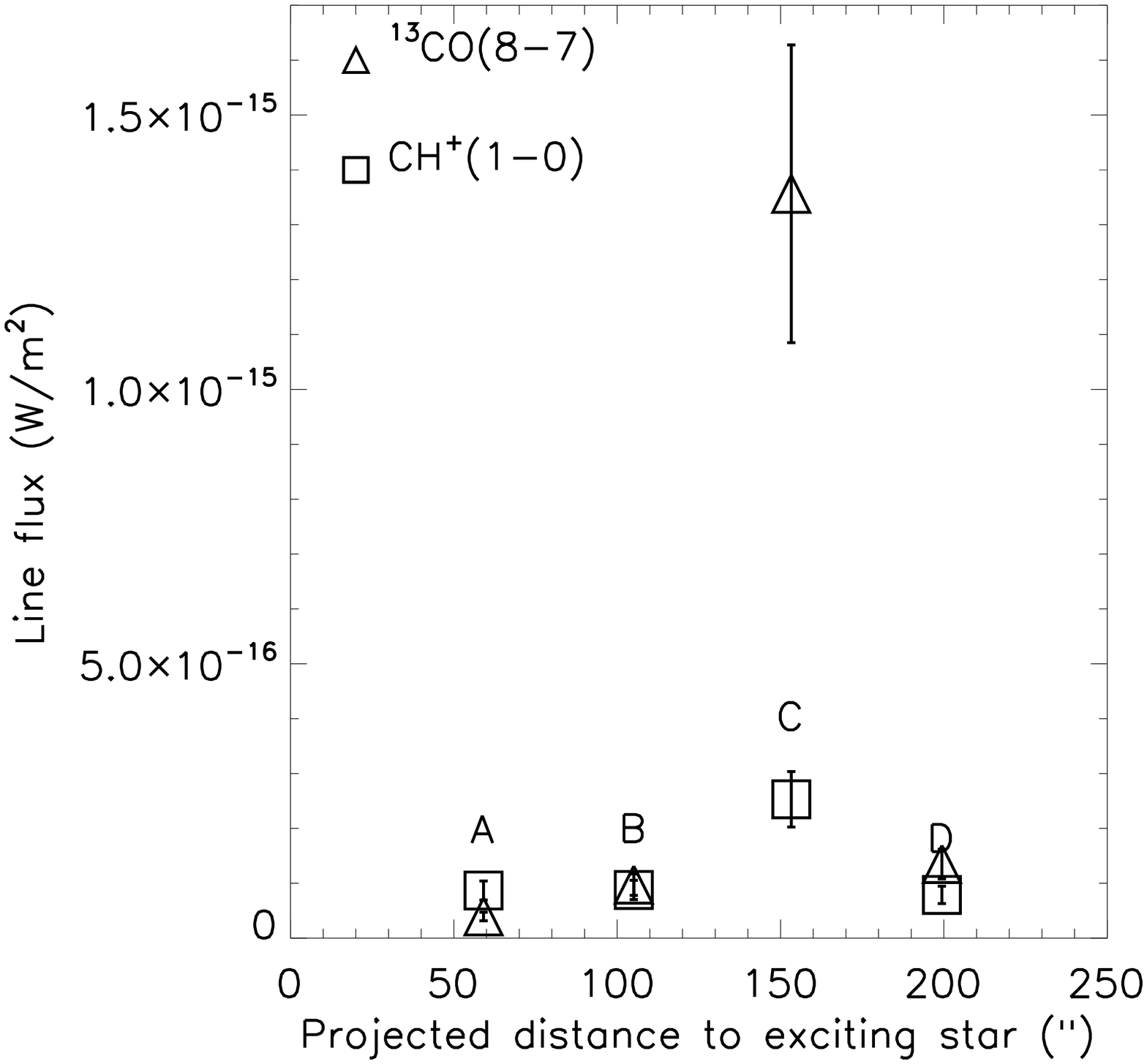}
\end{center}
\end{minipage}
\caption{Left panel: $^{12}$CH$^+$(1-0) continuum subtracted FTS spectra in the Orion bar region, overlayed on previous observations by \cite{Lis1998}, which combine dust and gas emission. The color map is the $\rm^{12}$CO(6-5) emission, the contours extending over the whole map are the continuum emission at 350$\mu$m and the contours displayed only for $\Delta$Ra above $-50$ correspond to $\rm^{13}$CO(6-5) emission. The Orion PDR exciting star is also shown and a cut passing through the bar discussed in the text. The stratification of the PDR appears clearly on Lis et. al. (1998) data, where the $^{12}$CO color map tracing hot gas is displaced toward the star with respect to $^{13}$CO contours tracing the dense molecular regions. Middle panel: same as CH$^+$ map for $\rm ^{13}$CO(8-7). Right panel: line fluxes along the cut presented in left panel for $\rm^{13}CO$ and $\rm CH^+$(1-0)}
\label{fig2}
\end{figure*}

%
The recorded spectra for $^{12}$CH$^+$ are overlaid on published maps by Lis et al. (1998) in Fig.~\ref{fig2}. The position of the O6 $\theta^1$ Ori C exciting star is also shown on the map. The expected geometry along the line of sight can be found in Fig.3 from \cite{Pellegrini2009} and in references therein. A cut through the expected best exposed projected line of sight is displayed (Fig.~\ref{fig2}, left panel) and the integrated fluxes along this cut are shown in the right panel.
 Estimates of the column density corresponding to the positions shown in Fig.~\ref{fig2}, which have been derived assuming thermal equilibrium with a mean Orion Bar temperature ($\rm\sim85\ K$), determined from ground based observations (Hogerheijde et al. 1995), are given in Table~\ref{table:3}. This table also presents abundance values for temperatures of 50 and 200 K, which probe different optical depths in the PDR. As noted above, CH$^+$ is highly reactive, easily destroyed by collisions and is not expected to be found in thermal equilibrium in the presence of strong UV pumping. Detailed analysis and interpretation of the measured fluxes is ongoing, but is beyond the scope of this letter.
%
%
\begin{table}[htbp]
\caption{CH$^+$ and CH  transitions in the SPIRE/FTS range}             
\label{table:2}      
\begin{center}                    
\tiny
\begin{tabular}{l c c c}        
\hline\hline                 
Freq (GHz) & Transition & A(s$^{-1}$)$\,^a$ & Ref \\    
\hline                        
CH$^+$ 	& $\rm J=1\leftarrow0$ &  &  \\      
835.079043(1.0)	&$1 - 0$	&5.96$\times$10$^{-3}$	&c\\
CH 	& $\rm N=1$ $\rm J=3/2\leftarrow1/2$ &  &  \\      
532.721333 (100)	&$\rm F1 = 1+ \leftarrow F2 = 1-$	&2.1$\times$10$^{-4}$     &b\\
532.723926 (40) 	&$\rm F1 =2+ \leftarrow F2 = 1-$	&6.2$\times$10$^{-4}$	&b\\
532.793309 (50)	&$\rm F1 =1+ \leftarrow F2 = 0-$	&4.1$\times$10$^{-4}$	&b\\
536.761145 (50)	&$\rm F1 =2- \leftarrow F2 = 1+$	&6.4$\times$10$^{-4}$	&b\\
536.781945 (50)	&$\rm F1 =1- \leftarrow F2 = 1+$	&2.1$\times$10$^{-4}$	&b\\
536.795678 (50)	&$\rm F1 =1- \leftarrow F2 = 0+$	&4.2$\times$10$^{-4}$	&b\\
CH 	& $\rm N=2\leftarrow1$ $\rm J=3/2\leftarrow3/2$ &  &  \\      
1470.6861(30) 	&$\rm F2 = 1- \leftarrow F1 = 2+$	&8.8$\times$10$^{-4}$	&d\\
1470.6890(30) 	&$\rm F2 =1- \leftarrow  F1 =1+$	&4.4$\times$10$^{-3}$	&d\\
1470.7356(30) 	&$\rm F2 =2- \leftarrow  F1 =2+$	&4.8$\times$10$^{-3}$	&d\\
1470.7385(30) 	&$\rm F2 =2- \leftarrow  F1 =1+$	&5.3$\times$10$^{-4}$	&d\\
1477.2901(30) 	&$\rm F2 =1+ \leftarrow  F1 =1-$	&4.4$\times$10$^{-3}$	&d\\
1477.3104(30) 	&$\rm F2 =1+ \leftarrow  F1 =2-$	&8.8$\times$10$^{-4}$	&d\\
1477.3630(30) 	&$\rm F2 =2+ \leftarrow  F1 =1-$	&5.3$\times$10$^{-4}$	&d\\
1477.3832(30) 	&$\rm F2 =2+ \leftarrow  F1 =2-$	&4.7$\times$10$^{-3}$	&d\\
\hline                                   
\end{tabular}
\end{center}
$^a$calculated using CDMS data parameters; $^b$\cite{Amano2000}; $^c$\cite{Pearson2006}; $^d$\cite{Brown1983}
\end{table}

\begin{figure}[h!]
\begin{center}
\includegraphics[width=0.75\columnwidth]{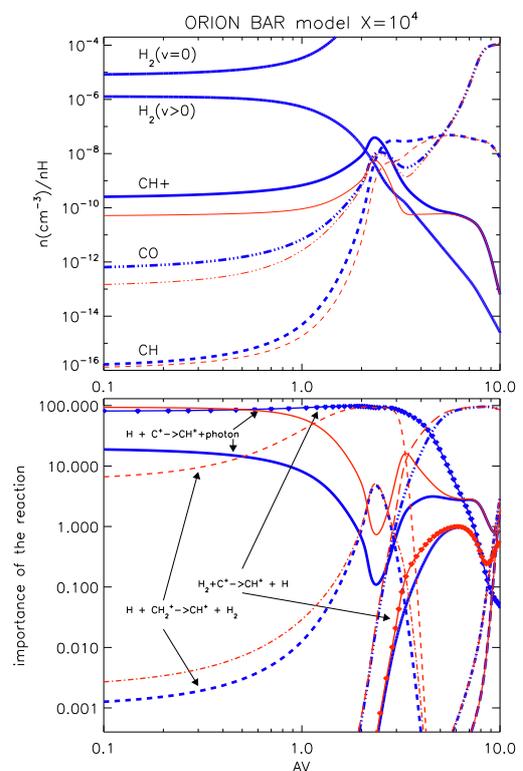}
\caption{Orion Bar isochore PDR model, using the Meudon code, as a function of the visual extinction normal to the front. The adopted PDR parameters are an incident flux of $\chi=10^4$ and a density of $\rm n_H=10^5 \mbox{ }cm^{-3}$. The upper panel shows model results in which the state by state vibrational excitation of H$_2$ in $\rm v\geq1$ (thick blue line) is considered as an internal energy involved in overcoming potential energy barriers, notably for the $\rm H_2+C^+$ reaction. The thin red line shows the results for the classical PDR model. The lower panel is a fractional representation of the relative importance of the methylidyne cation production reactions as a function of $\rm A_V$.}
\label{model}
\end{center}
\vspace*{-1.0cm}
\end{figure}

One of the striking features when comparing these plots is the prominence of CH$^+$ with respect to the $\rm^{13}CO(8-7)$ transition fluxes at the A and B positions, away from the Bar, probably corresponding to the surface of the cavity perpendicular to the line of sight. This tendency is reversed on the Bar (C position). From a preliminary analysis of $\rm^{13}CO$ SLW data along the A and B line of sight,  an excitation temperature of $\rm 85\ K^{+15}_{-45}$ is derived from the rotational diagram, corresponding to N$\rm_B$($^{13}$CO)$\rm=4.5-16.9\times10^{15} cm^{-2}$. We chose to focus on the B line of sight to avoid the effects of vignetting which are known to occur on the outer ring of bolometers. Such column densities would correspond to A$_V$ in the range $\rm 1-6$ at most,  and fractional abundances in the range $\rm 2-7\times10^{-10}$ for CH$^+$. We compare these estimates with a Meudon PDR code (Le Petit et al. 2006) model applied to the Orion Bar environment, described in more detail in \cite{Habart2010}. CH$^+$ abundances in A and B are only marginally compatible with a classical Orion PDR model (Fig. \ref{model}, red curves), whereas taking into account the vibrationally excited H$_2$ states (blue curves) where about 6\% of H$_2$ is in an excited vibrational state, enhances the formation rate by a factor of about 30 at low A$_V$.
The relative importance of the CH$^+$ production reactions, for both models, is shown in the lower panel of Fig. \ref{model}.
At low to moderate A$_V$, corresponding to the front and surface of the PDR, in the model using
only thermal rates, the radiative association reaction of H + C$^+$ dominates (blue curves). This reaction is, however, superseded by the H$_2$ + C$^+$ ion molecule reaction when the activation barrier can be overcome by the vibrational excitation of H$_2$ (red curves).
\begin{figure}[b]
\begin{center}
\includegraphics[width=0.75\columnwidth]{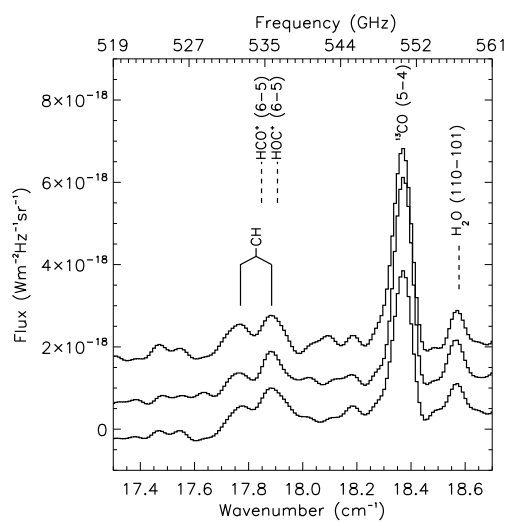}
\caption{Continuum subtracted spectra of the 3 positions on the Orion bar, displaying a doublet as expected for CH. }
\label{ch_lines}
\end{center}
\end{figure}

The fact that the CH$^+$ ion is seen extended all around the bar in the lower A$_V$ regions and in high abundance seems to support the importance of excited H$_2$ in the CH$^+$ formation route.
At lower frequencies, a pair of lines is observed at the position of the lambda-doublet CH transitions as shown in Fig.\ref{ch_lines}. In estimating the abundance of CH and CH$^+$ it has been assumed that the opacity is negligible with the species thermalised at the temperatures noted. Only the spectra on the Bar are shown. The signal-to-noise ratio of these two lines will improve as the SPIRE FTS data processing is improved and with deeper observations toward Orion. Higher signal-to-noise is required to ensure that potential contributions from HCO$^+$ and HOC$^+$ line emission to this spectral region are considered, particularly for the higher frequency component.

\begin{table}
\caption{Column densities along the cut}             
\label{table:3}      
\begin{center}                    
\begin{tabular}{l l l c}        
\hline\hline                 
Position & $^{13}$CO ($\rm 10^{15} cm^{-2}$) & CH$^+$($\rm 10^{12} cm^{-2}$) & CH($\rm 10^{12} cm^{-2}$) \\    
\hline                        
A & $\rm1.2(85K)^{1.3(200K)}_{5.6(50K)}$ 		& $\rm2.2(85K)^{3.8(200K)}_{1.9(50K)}$	&  \\    
B & $\rm4.8(85K)^{3.1(200K)}_{13.8(50K)}$ 		& $\rm2.2(85K)^{3.8(200K)}_{1.9(50K)}$	&  \\    
C & $\rm67.4(85K)^{43.5(200K)}_{191.7(50K)}$ 	& $\rm6.4(85K)^{11(200K)}_{5.5(50K)}$	& $\rm\leq4.5(85K)^{8.4(200K)}_{3.6(50K)}$ \\    
D & $\rm6.7(85K)^{4.3(200K)}_{19.1(50K)}$ 		& $\rm2.0(85K)^{3.4(200K)}_{1.7(50K)}$	&  \\    
\hline                                   
\end{tabular}
\end{center}
The angular size assumes a filled gaussian beam of 36" at CO, CH$^+$ transition frequencies and 33.9" at CH ones.
\vspace*{-12pt}
\end{table}

\subsection*{G29.96-0.02 and G32.80+0.19}
\begin{figure*}[t]
\begin{minipage}{0.78\columnwidth}
\begin{center}
\hspace*{-0.5cm}
\includegraphics[width=0.85\columnwidth]{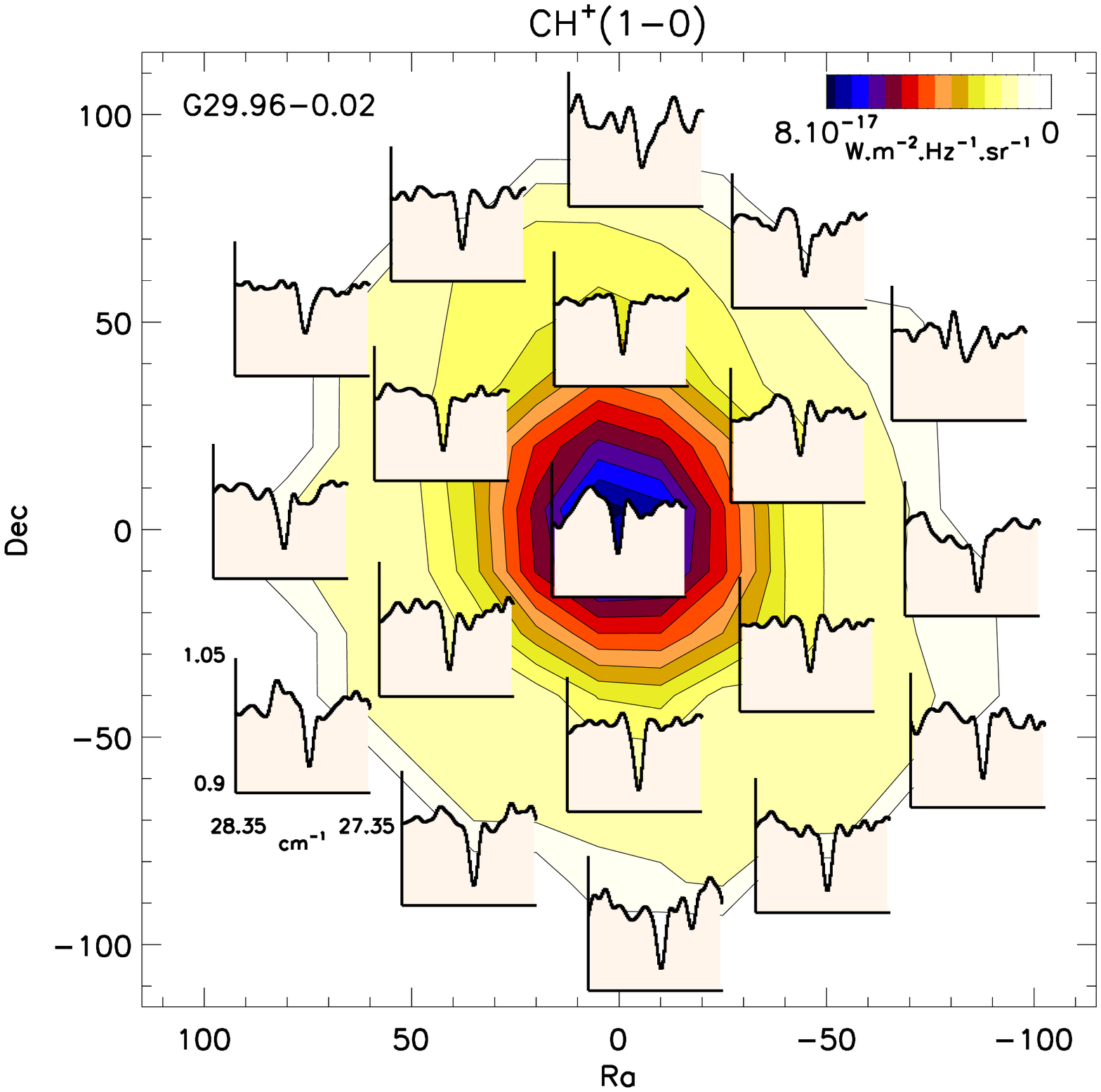}
\end{center}
\end{minipage}
\begin{minipage}{0.45\columnwidth}
\begin{center}
\hspace*{0.5cm}
\caption{Left panel: G29.96-0.02 CH$^+$(1-0) continuum divided transmittance spectra, overlaid on the continuum intensity map at the same frequencies, extending away from the ultra compact HII source. Right panel:  same for the G32.80+0.19 HII region. The outer detectors are vignetted so have been omitted.}
\label{fig5}
\end{center}
\end{minipage}
\begin{minipage}{0.78\columnwidth}
\begin{center}
\includegraphics[width=0.85\columnwidth]{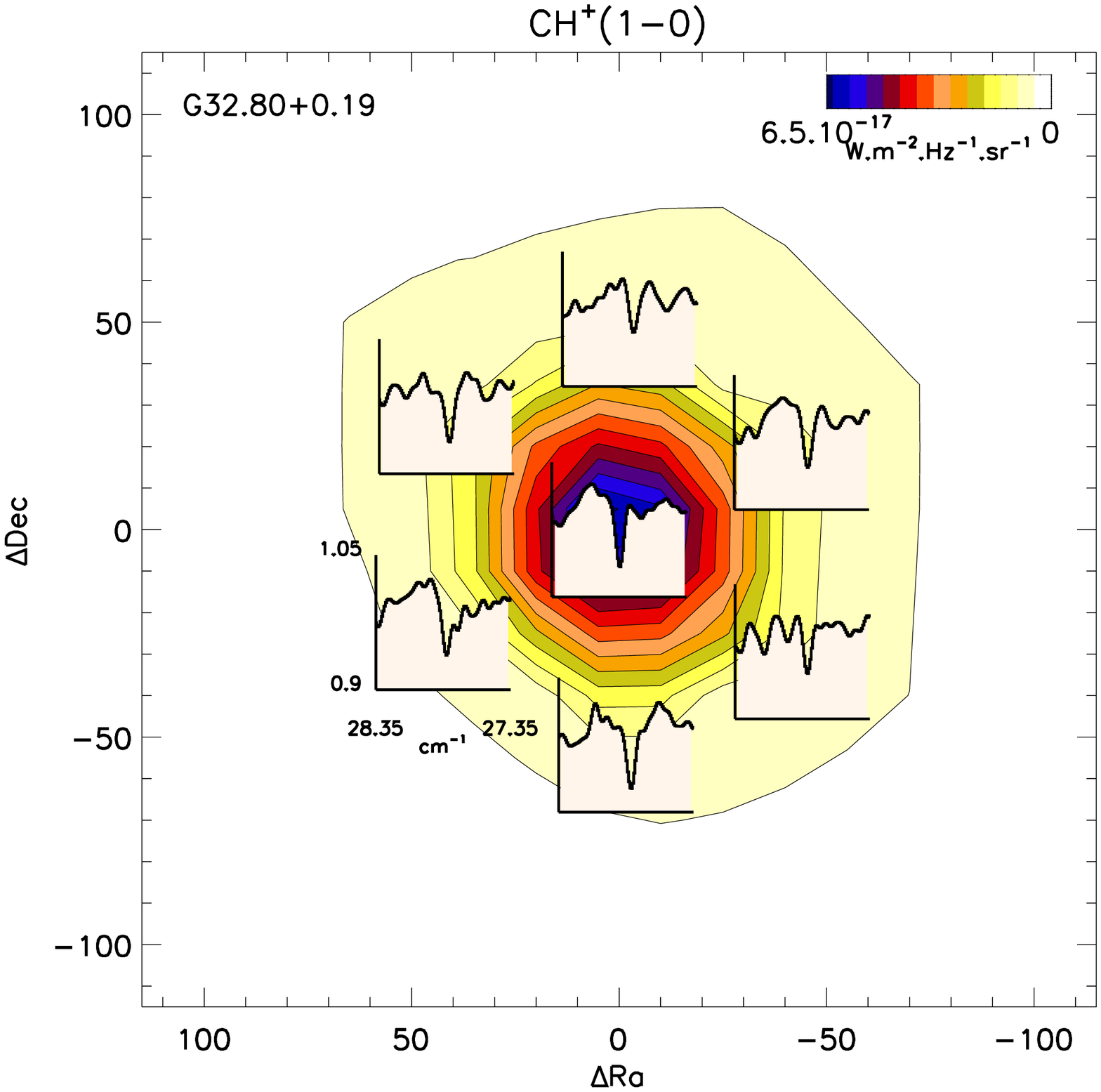}
\end{center}
\end{minipage}
\vspace*{-0.5cm}
\end{figure*}

G29.96-0.02 and G32.80+0.19 are two ultra compact HII regions (e.g. Wood and Churchwell 1989), located near the Galactic plane. A global analysis of G29.96-0.02 SPIRE/FTS data can be found in this issue (Kirk et al. 2010). Both HII regions display a CH$^+$ absorption line at a few percent of the continuum level.
From a study of the line-to-continuum ratio maps shown in Fig.\ref{fig5}, it appears that for each continuum level across the maps, the amount of absorption remains essentially the same, within the signal-to-noise limits for the outer pixels. This suggests that the methylidyne ion is, as expected from its electronic visible absorption transitions, present in the foreground diffuse interstellar medium in relative high abundances. The apodized resolution of the spectra is 0.0724 cm$^{-1}$, corresponding to about 780~km s$^{-1}$
for the considered CH$^+$ transition. Although the absorptions appear shallow, at the SPIRE/FTS resolution, a 5\% absorption corresponds to a saturated line about 38~km s$^{-1}$ wide. This is consistent with the velocity dispersion of the foreground ISM as seen in HI absorption (e.g. Fish et al. 2003, Fig.1). This precludes us from deriving stringent column densities at this level of analysis.

In subsequent observations with the SPIRE FTS the CH$^+$ line has also been measured in the planetary nebula NGC7027 (Wesson et al. 2010) and in the ultra luminous infrared galaxy MRK231 (Van der Werf et al. 2010). The SPIRE FTS promises to yield a harvest of methylidynium data from a wide variety of sources.
\section{Conclusions}

We have presented the first detection of the lowest rotational transition of the methylidyne cation CH$^+$ in an astronomical source.
We have also shown strong evidence for the detection of the lambda doublet emission from the CH molecule. The detection of CH$^+$ in a variety of environments shows that spectral mapping of this and other species will provide a powerful probe of the physics of the interstellar medium. These results have only been made possible by the sensitivity and broad spectral coverage of the SPIRE FTS.

\begin{acknowledgements}
SPIRE has been developed by a consortium of institutes led by Cardiff Univ. (UK) and including Univ. Lethbridge (Canada); NAOC (China); CEA, LAM (France); IFSI, Univ. Padua (Italy); IAC (Spain); Stockholm Observatory (Sweden); Imperial College London, RAL, UCL-MSSL, UKATC, Univ. Sussex (UK); Caltech, JPL, NHSC, Univ. Colorado (USA). This development has been supported by national funding agencies: CSA (Canada); NAOC (China); CEA, CNES, CNRS (France); ASI (Italy); MCINN (Spain); SNSB (Sweden); STFC (UK); and NASA (USA). DAN acknowledges support from NSERC. The authors thank Jacques LeBourlot and Guillaume Pineau des For\^ets for their assistance with the analysis/modeling of the Orion Bar.
\end{acknowledgements}

\end{document}